\documentclass[a4paper]{jpconf}
\usepackage{graphicx}
\usepackage[rightcaption]{sidecap}
\usepackage{cite}
\begin{document}
\title{Investigation of the quasifission process by theoretical analysis of experimental data \\
of fissionlike reaction products}

\author{G Giardina$^{1,2}$, A K Nasirov$^{3,4}$, G Mandaglio$^{1,2}$, F Curciarello$^{1,2}$, V~De~Leo $^{1,2}$, G Fazio$^{1,2}$, M Manganaro$^{1,2}$, M Romaniuk$^{1,2}$,  C Sacc\'a$^5$},

\address{$^1$ Dipartimento di Fisica, Universit\`a di Messina, I-98166 Messina, Italy\\
  $^2$  Istituto Nazionale di Fisica Nucleare, Sezione  di Catania, I-95123 Catania,  Italy\\
   $^3$Joint Institute for Nuclear Research, 141980, Dubna, Russia \\
   $^4$ Institute of Nuclear Physics, 100214,  Tashkent, Uzbekistan\\
$^5$ Dipartimento di Scienze della Terra, Universit\`a  di Messina, I-98166 Messina, Italy}

\ead{giardina@nucleo.unime.it}

\begin{abstract}
The fusion excitation function is the important quantity in
planning experiments for the synthesis of superheavy elements. 
Its values seem to be determined by the
experimental study of the hindrance to complete fusion by the
observation of  mass, angular and energy distributions of the
fissionlike fragments. There is ambiguity in establishment of the
reaction mechanism leading to the observed binary fissionlike
fragments. The fissionlike fragments can be produced in the
quasifission, fast fission, and fusion-fission processes which
have overlapping in the mass (angular, kinetic energy)
distributions of fragments. The branching ratio between
quasifission and complete fusion strongly depends on the
characteristics of the entrance channel. In this paper we consider
a wide set of reactions (with different mass asymmetry and mass
symmetry parameters) with the aim to explain the role played by
many quantities on the reaction mechanisms. We also present the
results of study of the $^{48}$Ca+$^{249}$Bk reaction used to
synthesize superheavy nuclei with Z = 117 by the determination of
the evaporation residue cross sections and the effective fission
barriers $<B_{\rm f}>$  of excited nuclei formed along the
de-excitation cascade of the compound nucleus.
\end{abstract}

\section{Introduction}
\label{intro}
The experimental and theoretical investigations of reaction dynamics connected with the formation of composed system is nowadays the main subject of the nuclear reactions.
At the first stage of reaction of heavy ion collisions  the full momentum transfer can occur (this event is defined as capture)  if there is a well in the nucleus-nucleus
potential in dependence on the values of  relative kinetic energy and friction coefficients \cite{epja192004,prc812010}. At capture,
the two reacting nuclei form a rotating nuclear system at near Coulomb barrier energies. During its evolution this system can be transformed into compound nucleus or
it re-separates into two fragments which may differ from the initial nuclei in the entrance channel \cite{epja82000,npa671,jpsj72}.   During the evolution of DNS its two
nuclei  may change their masses  $A_1$, $A_2$ and charges $Z_1$, $Z_2$ but with constant total mass $A=A_1+A_2$ and charge $Z=Z_1+Z_2$. The DNS should overcome the
intrinsic fusion barrier $B^*_{\rm fus}$  (it is equal to the difference between  the maximum  value of the driving potential and its value corresponding to the
initial charge asymmetry) to reach  the compound nucleus state  through more mass asymmetric configurations. The intense of the break up of DNS into two nuclei
(quasifission channel)  in competition with the complete fusion  is characterized  by the value of the quasifission barrier $B_{\rm qf}$ (the depth of the pocket
in the nucleus-nucleus potential) \cite{epja192004,prc812010}.
The mass asymmetry parameter of quasifission fragments may be larger or smaller than that of nuclei in the entrance channel. It is well known the case of the
quasifission with small mass asymmetry in comparison with one of initial nuclei.
In this case the maximum of the mass distribution of quasifission fragments places between the masses of projectile (or target)  and symmetric fission fragments.

On the other hand, due to the exchange with neutrons and protons between the nuclei constituting the DNS the mass asymmetry parameter increases being larger
than the mass asymmetry in the entrance channel (evolution to complete fusion
direction) and at the same time there is also a relevant probability of DNS to decay into two nuclei. The decay probability depends on the excitation energy  and quasifission barrier
$B_{\rm qf}$ of DNS at a given configuration. Of course, the mass distribution of quasifission products are related to the characteristics of the entrance channel (beam energy
and orientation angles of the axial symmetry axis of reacting nuclei).

The capture events surviving quasifission populate the complete fusion channel. Consequently, the deformed mononucleus may reach the fully equilibrated statistical shape of the
compound nucleus (CN), or if there is no the fission barrier $B_{\rm f}$ the system immediately  decays into two fragments (fast fission process). The latter phenomenon occurs
only  at high angular momentum $\ell$ for which the fission barrier of the complete fusion system disappears  ($B_{\rm f}(\ell>\ell_{\rm f})$= 0).
Therefore, the fast fission process takes place only in the reactions at high angular momentum values ($\ell>\ell_{\rm f}$) while the quasifission process takes place at all values of $\ell$ contributing to the capture reaction.

Finally, in the last stage of  nuclear reaction, the formed CN may de-excite by  emission of light particles or undergoes to fission (producing fusion-fission fragments).
The reaction products that survive fission are the evaporation residues (ER)\cite{epja222004,prc722005}.
The registration of ER is clear evidence of the compound nucleus formation, but generally the determination of ER's only  it is not enough to determine the complete fusion
cross section and to understand the dynamics  of the de-excitation cascade of CN if the fission fragments are not included into consideration.  On the other hands, it is
difficult for sure correct determination of the fusion-fission rate in the cases of  overlapping of the mass and angle distributions of the products of different processes
(quasifission, fast fission and fusion-fission) because sorting out the registered fissionlike fragments according to the mechanism of its origin is connected with some assumptions.

Moreover, by observing the fission products of fissile nuclei formed in the
 in neutron (or very light particles)-induced reactions  with the heavy targets one can conclude that the low excited  compound nucleus (at about $E_{\rm CN}^* <$ 10 MeV)
 decays into very asymmetric fission fragments (near to the shell closure), while the compound nuclei formed in heavy ion collisions at intermediate or high excitation energy
 ($E_{\rm CN}^* >$ 20 MeV) undergo fission forming the mass symmetric fragments.
Starting from these general observations some researchers put forward the idea that the complete fusion process of two colliding nuclei may be considered as the inverse process
to fission. The   authors of the papers \cite{prc782008,prc792009}  argued that since the fission of a compound nucleus in heavy ion collisions produces just symmetric fragments,
 then in the collisions of two symmetric (or almost symmetric) nuclei  complete fusion has to be a very probable process. But, unfortunately this is not true.  For  systems of
 colliding nuclei heavier than $^{110}$Pd+$^{110}$Pd the fusion does not occur absolutely, while for the reactions like $^{100}$Mo+$^{100}$Mo, $^{96}$Zr+$^{96}$Zr, $^{96}$Zr+$^{100}$Mo,
  $^{100}$Mo+$^{110}$Pa or induced by projectiles higher than Zn, Ge, Kr  there is a strong hindrance to fusion.

Following  the previous reasons one can affirm that the hypothetical $^{132}$Sn+$^{120}$Cd reaction should  lead to the $^{252}$Cf CN  since  $^{120}$Cd (with $Z$=48 near the shell
 closure 50) and $^{132}$Sn (with double shell closure $Z$=50 and $N$=82) are produced with highest yields in spontaneous fission of $^{252}$Cf. But our estimation for this reaction
 does not give meaningful fusion probability ($P_{\rm CN}<5\times 10 ^{-7}$).

The simple reason resides in the peculiarities of the reaction dynamics. In the spontaneous fission of $^{252}$Cf the average value of angular momentum distribution of the fragments is close to zero, but if we want to reach the  $^{252}$Cf compound nucleus,  by the hypothetical $^{132}$Sn+$^{120}$Cd reaction (or by the realistic $^{132}$Sn+$^{116}$Cd reaction leading to the  $^{248}$Cf CN), the average value of angular momentum distribution of DNS in the entrance channel may be about $<\ell>=50 \hbar$ or higher by increasing the beam energy. The $<\ell>$ value is calculated  as

\begin{eqnarray}\label{uno}
{<\ell\, (E_{\rm c.m.};\,\alpha_{\rm P},\alpha_{\rm T})>}&=& \frac{\sum_{\ell=0}^{\ell=\ell_d}
\ell\, <\sigma^{(\ell)}_{(cap)}>_{\alpha_{P},\alpha_{T}}\,(E_{\rm c.m.})}
{\sum_{\ell=0}^{\ell=\ell_d}
<\sigma^{(\ell)}_{(cap)}>_{\alpha_{P},\alpha_{T}}(E_{\rm c.m.})}\,\,\,\,\, .
\end{eqnarray}

In the  fusion reaction mechanism, at the first stage, DNS which is formed after capture of projectile by
the target nuclei should survive quasifission (re-separation of nuclei of DNS). Due to the hindrance connected with quasifission for these system the fusion probability $P_{\rm CN}$
 should be lower than 10$^{-7}$ for
 small values of angular momentum. This low probability of complete fusion becomes more lower for the excited and fast rotating deformed mononucleus which undergoes fast fission
before the system can reach the compact shape of compound nucleus.

Also in the cases of the explored $^{22}$Ne+$^{250}$Cf (more mass asymmetric  system), $^{24}$Mg+$^{248}$Cm, $^{28}$Si+$^{244}$Pu, $^{34}$S+$^{238}$U, and $^{40}$Ar+$^{232}$Th
(less mass asymmetric  system) reactions, the compound nucleus $^{272}$Hs  is formed with different angular momentum distributions   in dependence on entrance channels even the
defined excitation energy $E_{\rm CN}^*$ has been achieved. Therefore, such compound nuclei decay by different yields of reaction products. The fusion-fission fragment mass
distributions are peaked at around the $^{136}$Xe nucleus with different dispersions and average angular momentum distributions in connection with the various entrance channels.
If we calculate the formation probability of the $^{272}$Hs  compound nucleus in  the mass symmetric $^{136}$Xe+$^{136}$Xe reaction at the same fixed excitation energy $E^*_{\rm CN}$
as in the considered $^{22}$Ne+$^{250}$Cf reaction (where $P_{\rm CN}\simeq$ 1), we do not meaningfully  reach such a compound nucleus($P_{\rm CN}<$10$^{-10}$).  The angular momentum
distribution for the $^{136}$Xe+$^{136}$Xe  collision at the capture stage is completely different and all conditions of reaction dynamics  lead to deep inelastic and quasifission products.

In this context, for  the $^{136}$Xe+$^{132}$Sn  ($P_{\rm CN}<$10$^{-8}$) and $^{132}$Sn+$^{176}$Yb  ($P_{\rm CN}<$5$\times$10$^{-11}$) reactions, one can observe the same above-described hindrance to  complete fusion.

\section{Capture and Deep Inelastic Collision }

Our theoretical capture cross section  $\sigma_{\rm cap}$ includes all damped reactions, excluding deep inelastic collisions (DIC). The partial capture cross sections are contributed
by the full momentum transfer events corresponding to the trapping into a pocket in the nucleus-nucleus potential    after dissipation of the relative kinetic energy.  Differently,
the DIC  events are not characterized by the full momentum transfer and collision paths are not trapped in the pocket of the nucleus-nucleus potential. The DIC events  are not connected with any pocket of
a potential; for example at large values of $\ell$ there is no pocket but the DIC events take place.  The partial capture cross section is calculated by the following formula 

 \begin{equation}
 \label{parcap}
 \sigma^{\ell}_{cap}(E_{\rm c.m.})=
 \pi{\lambda\hspace*{-0.23cm}-}^2
{\cal P}_{cap}^{\ell}(E_{\rm c.m.})
 \end{equation}

with the capture probability ${\cal P}_{cap}^{\ell}$ which is found by the solution of the motion equations \cite{mpla202005,jpsj772008}.

\begin{figure}[tbp]
\vspace*{-2.5cm}
\par
\begin{center}
\resizebox{1.0\textwidth}{!}{\includegraphics{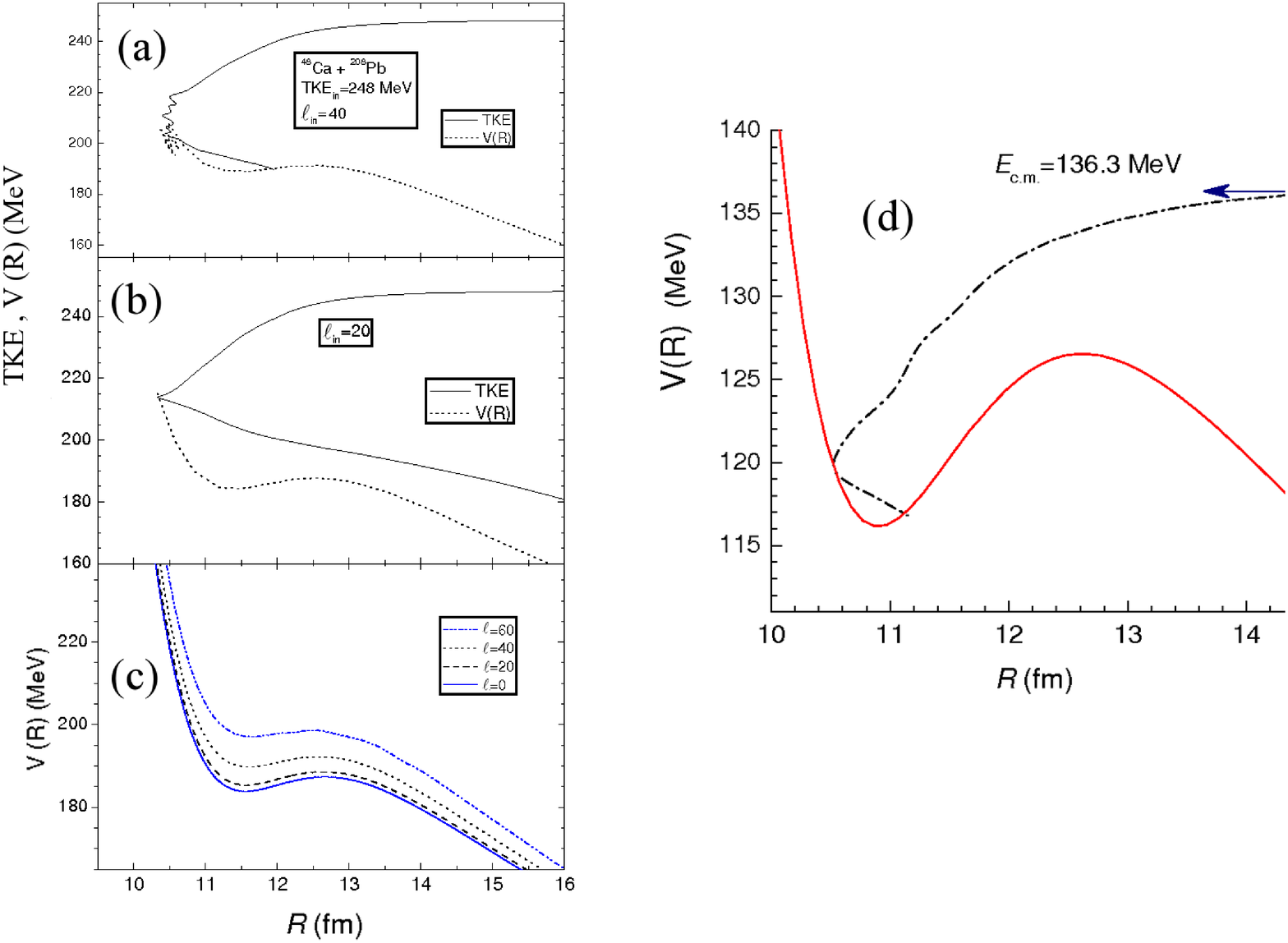}}
%Fig3
% Here is how to import EPS art
\vspace*{-1.5cm}
\end{center}
\caption{Capture (a) and deep inelastic collision  (b) for the
$^{48}$Ca+$^{208}$Pb reaction. Dependence of nucleus-nucleus
potential (c) on the orbital angular momentum $\ell$. Illustration of capture path (d) (dot dashed line) into potential well
(solid line) as obtained by the numerical solution of the equation of relative motion of
colliding nuclei with the initial energy $E_{\rm c.m.}=136.3$ MeV and $\ell$=0 for
the $^{32}$S+$^{184}$W reaction\cite{prc812010}.}\vspace*{-0.3cm}
\label{prima}
\end{figure}

\begin{figure}[h]
\vspace*{0.3cm}
\par
\begin{center}
\resizebox{0.8\textwidth}{!}{\includegraphics{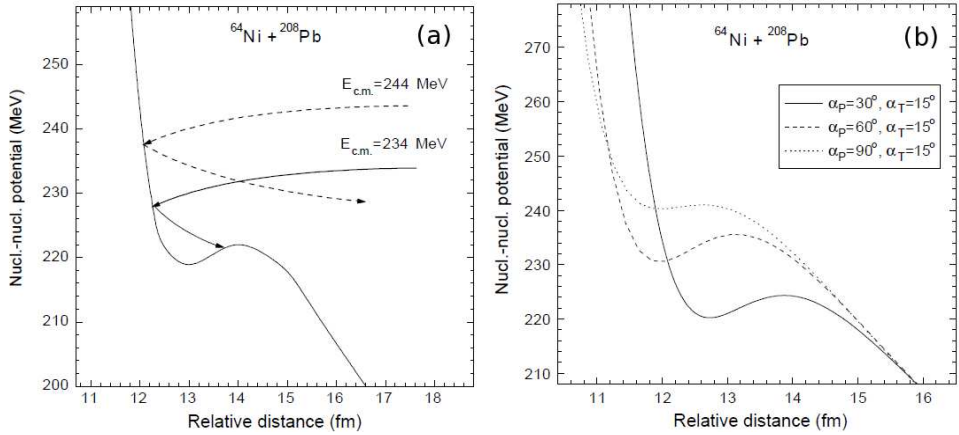}}
%Fig3
% Here is how to import EPS art
\vspace*{-0.7cm}
\end{center}
\caption{(a) Dependence of capture (solid curve) and deep inelastic collision
(dashed curve) processes for nucleus-nucleus collisions on the
internuclear potential $V$($R$) and on the initial values of the
beam energy $E_{\rm c.m.}$.(b) Dependence of the nucleus-nucleus potential on the
mutual orientation of colliding nuclei; solid line: $\alpha_1$ = 30$^\circ$
and $\alpha_2$= 15$^\circ$; dashed line:  $\alpha_1$ = 60$^\circ$
and $\alpha_2$= 15$^\circ$; dotted line:
$\alpha_1$= 90$^\circ$
and $\alpha_2$= 15$^\circ$.}
\label{seconda}\vspace*{-0.5cm}
\end{figure}

In the calculation of nucleus-nucleus potential, which includes the Coulombian $V_{\rm Coul}$($Z$,~$A$,~$R$), nuclear $V_{\rm n}$($Z$,~$A$,~$R$) and rotational
$V_{\rm rot}$($Z$,~$A$,~$R$,~$\ell$) parts, we take into account the static and dynamic deformations,
and orientation angles of the axial symmetry axes of reacting nuclei, at initial stage.

 \begin{SCfigure}[][h]
\centering
\includegraphics[width=0.75\textwidth]{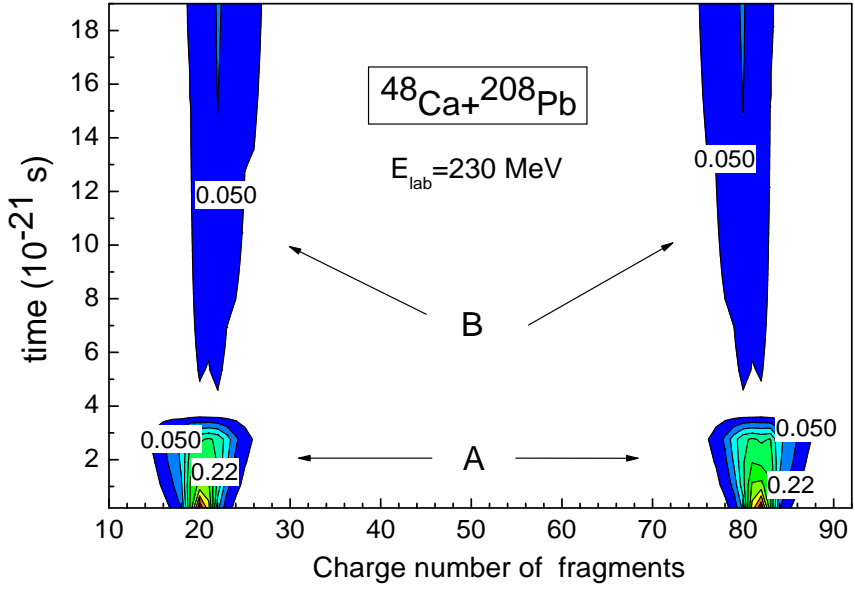}
\vspace*{-6.cm}
\caption{Time dependence of the theoretical results for the mass
distribution of the deep inelastic collision (A) and quasifission
(B) fragments in the $^{48}$Ca+$^{208}$Pb reaction.}\label{fig4mpla}
\end{SCfigure}

In Fig. \ref{seconda} we present an example of capture events with different potential wells and its dependencies on the angular momentum and orientation angles of axes of reacting nuclei;
moreover, in figures  we also present the cases of capture and DIC events by changing  the beam energies and angular momentum values.

 The lifetime of an excited DNS for a given reaction depends on the initial collision energy $E_{\rm c.m.}$ and angular momentum distribution values (see formulae (6)-(8) and
 Fig.4 of our paper\cite{{ijmpe182009}}).
 The mass and charge, as well as shapes of nuclei constituting the DNS are changed during interaction time. The evolution of DNS depends on its excitation energy, orientation
 angles of the axial symmetry axes  and shell structures of reacting nuclei. During its evolution DNS can evolve to complete fusion  or can decay into two fragments (quasifission process).
 The competition between these two processes is related to the values of intrinsic fusion barrier $B^*_{\rm fus}$ and quasifission barrier   $B_{\rm qf}$\cite{epja222004,prc812010,prc722005}
 depending on the peculiarities of reacting nuclei, beam energy and angular momentum distribution.
\begin{figure}[tbp]
\vspace*{1.9cm}
\par
\begin{center}
\resizebox{1.0\textwidth}{!}{\includegraphics{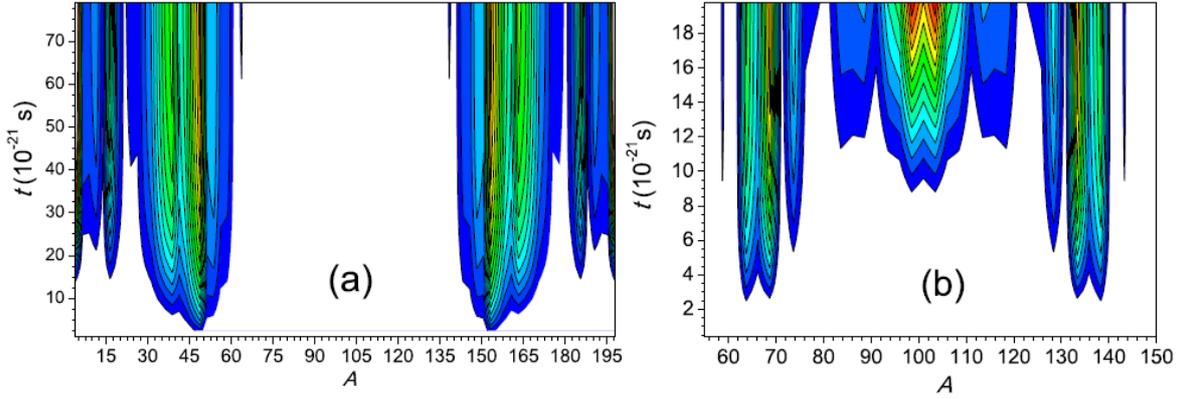}}
%Fig3
% Here is how to import EPS art
\vspace*{-1.2cm}
\end{center}
\caption{ The mass distribution of the quasifission product yields obtained in the $^{48}$Ca+$^{154}$Sm reaction at $E_{\rm c.m.}$=140 MeV (panel (a)) and  $E_{\rm c.m.}$=160 MeV (panel (b)) as a function of the lifetime of the dinuclear system formed at capture stage. }
\label{figjpcs2}
\end{figure}

In many cases and conditions the mass and angular momentum
distributions of quasifission fragments  can overlap with the mass
and angular distributions of fusion-fission fragments, or/and with
the ones of the fast fission
fragments\cite{mpla202005,prc792009n}, leading to the real
difficulties in the experimental analysis in order to sort out the
true yields of fragments belonged  to various reaction mechanisms
(see for example \cite{mpla202005,plb2010} and Fig. 3 of Ref.
\cite{jpcs2010}).

\begin{figure}[tbp]
\vspace*{-1.5cm}
\par
\begin{center}
\resizebox{0.8\textwidth}{!}{\includegraphics{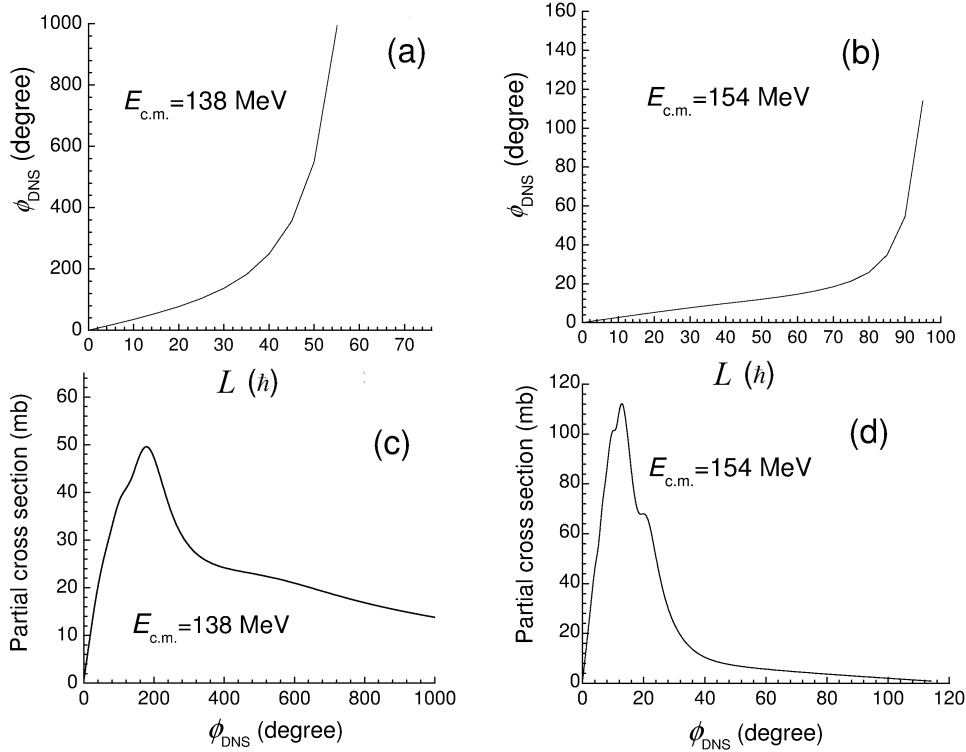}}
%Fig3
% Here is how to import EPS art
\vspace*{-0.2cm}
\end{center}
\caption{The rotation angle $\phi_{DNS}$ of the dinuclear system forming the $^{58}$Cr and $^{144}$Ce fragments (reached by the $^{48}$Ca+$^{154}$Sm reaction at two different beam energies) as a function of the orbital angular momentum $L$ in (a) and (b), and the angular distributions of the  quasifission  $^{58}$Cr and $^{144}$Ce fragments at decay of DNS in (c) and (d).}\vspace*{-0.2cm}
\label{figjpcs3}
\end{figure}

Usually, in reactions with actinides the maximum of the yield  of
the quasifission products is observed in the mass region between
masses of the projectile-like products and symmetric masses. In
the $^{48}$Ca + $^{208}$Pb reaction, the maximum of the
quasifission products is very close  to the region of
projectilelike and targetlike fragments (see Fig. 3 of Ref.
\cite{mpla202005}, as well as Fig. 1 of Ref. \cite{Itkis02G} and
Fig. 8 of Ref. \cite{Bock}). This is due to the fact that
projectile and target are double magic nuclei. As a result the
mass distributions of the capture, {\it i.e.} quasifission
reactions, and deep inelastic collisions are mixed. The reaction
time of the deep inelastic collisions is much smaller than that of
quasifission and fast fission reactions, because the last two
reactions take place if  capture occurs, {\it i.e.} projectile has
been trapped into potential well (see Fig. \ref{prima}).

It is difficult to separate without uncertainty products of these two processes in experimental data. To demonstrate the importance of this circumstance, the
mass distributions of the deep-inelastic and capture reactions
were calculated by the method developed on the basis of the
DNS concept \cite{Jolos86,Nasi02}. The results for the $^{48}$Ca+$^{248}$Pb reaction are
presented in Fig. \ref{fig4mpla}. The reason of this result can be seen from the driving potential
which has a deep valley around initial charge asymmetry due to the
nuclear shell effects. Therefore, the sufficient part of the
yields of quasifission and deep-inelastic products are mixed in
the mass and charge distributions. The difference between them is in
their angular distributions which are connected with the lifetime
of the formed rotated DNS. The overlap  of the
maximum of the mass distribution of quasifission with that of the
deep  inelastic collisions can lead to conclusion that there is  no  the
contribution of the quasifission process in the total reaction
cross section. This explains the strong deviation of the
theoretical values of the capture cross section at low energies
from the experimental capture cross section  which was determined
by the symmetric fragmentation only.

So, the  mixing of the mass distributions of the deep inelastic
collisions  and quasifission processes causes an ambiguous
determination of the capture cross section from the experimental
mass distributions in case of reactions with magic or double magic
nuclei, like to the case of the reaction under discussion. As well
as the mixing of the mass distributions of the quasifission, fast
fission and fusion-fission reactions makes difficult the
estimation of the complete fusion cross section leading to the
genuine compound nucleus. Due to the strong dependence of the mass
distribution on the peculiarities of nuclear shell structure, the
proportion of mixing is changed from one reaction to another
reaction. The theoretical study of this problem is useful to
establish the reactions and beam energies which are favorable to
obtain the maximum cross section of formation of the compound
nucleus with the relatively low angular momentum and excitation
energy.

 We can state that, at the low energies,  the products of the
 symmetric fragmentation in the $^{48}$Ca+$^{208}$Pb reaction
are generally formed at usual fission of
CN, and therefore they are related to the fusion cross section. At
energies $E_{lab}>212$ MeV the contribution of the fast fission
appears and it increases by the beam energy. That will be shown in Fig. \ref{fasfis}.

Moreover, Fig. \ref{figjpcs2} shows the mass distribution of the quasifission  products of the $^{48}$Ca+$^{154}$Sm reaction as a function of the DNS lifetime, following the capture reaction at two $E_{\rm c.m.}$ energies of 140 MeV (panel (a)) and 160 MeV (panel (b)). At lower energies (panel (a)) the mass distribution of quasifission fragments are asymmetric;   at higher beam energy (panel (b)), also appear mass symmetric quasifission fragments. The latter mass distribution of quasifission fragments overlap to the fusion-fission mass distribution leading to a relevant uncertainty in experimental determination of the fusion cross section.

In Fig. \ref{figjpcs3} we present the results of the  $^{48}$Ca+$^{154}$Sm reaction regarding the determination of the rotational angle $\phi_{DNS}$ of the dinuclear system  at two $E_{\rm c.m.}$ beam  energies of 138 and 154 MeV (see panels (a) and (b), respectively) versus the angular momentum $L$ values of DNS, and the angular distributions of the quasifission $^{58}$Cr and $^{144}$Ce fragments at decay of DNS for the energies $E_{\rm c.m.}$=138 and 154 MeV, (see panels (c) and (d), respectively). As one can see,  at $E_{\rm c.m.}$=138 MeV the decay of DNS into the two mentioned fragments lead to large rotational angle $\phi_{DNS}$ at intermediate and high angular momentum components (panel (a)) as well as a wide  angular distributions of the quasifission $^{58}$Cr and $^{144}$Ce fragments at all DNS rotational angles (panel (c)); while, at high energy $E_{\rm c.m.}$=154 MeV, the rotational angle $\phi_{DNS}$ is small up to large angular momentum values (panel (b)), and the angular distribution of the quasifission fragments is strongly peaked to low rotational angle values of DNS (panel(d)). This means that at $E_{\rm c.m.}$=138 MeV for $L >$ 35$\hbar$ the rotation angle is enough large and the angular distribution of the quasifission fragments may be nearly isotropic.
The maximum of the angular distribution moves to small angles at larger energies:  $E_{\rm c.m.}$=154 MeV the most part of fragments is distributed to the forward angles and to backward angles for the relative fragment-partner (see panel (d)), and the maximum of the angular distribution are concentrated around $\phi_{DNS}$ = 15$^\circ$ in the center-of-mass system. Therefore the large part of quasifission fragments passes beside the nearest to the beam detector. This phenomenon caused the conclusions of the authors of Ref. \cite{galina07} about decreasing of quasifission events at energies $E_{\rm c.m.}>$154 MeV. The reason of decreasing in the experimental events of quasifission at low energies $E_{\rm c.m.}<$138 MeV is the isotropic angular distribution of some part of quasifission fragments which were considered as the compound nucleus fission fragments.

\section{Evolution of DNS : competition between quasifission and complete fusion processes}

The composite system formed at capture stage evolves by exchanging
nucleons between the two nuclei constituting the DNS. During its
evolution the DNS can reach the shape of a deformed mononucleus
(complete fusion) or it can break up into two fragments
(quasifission) without   reaching the complete fusion stage. In
the first case the nuclear system has to reach the statistical
equilibrate shape of the CN with the angular momentum $\ell_{\rm
CN}$. The equilibrium shape of compound nucleus depends on the
value of $\ell_{\rm CN}$ which determines the fission barrier
$B_{\rm f}$ providing its stability against to fission. It is well
known that $B_{\rm f}$=0 if the value of $\ell_{\rm CN}$ is larger
than its critical value $\ell >\ell_{\rm f}$ for the compound
nucleus with the given mass and charge number \cite{Sierk86}.
Therefore, the events of complete fusion  for $\ell >\ell_{\rm f}$
cannot reach the equilibrium shape of CN because the mononucleus
immediately decays into two fragments.  This fissionlike process
is called fast fission because compound nucleus is not formed.
Competition between complete fusion and quasifission occurs at all
values of the orbital angular momentum. Therefore, the partial
capture cross section is contributed by the following terms:
 \begin{equation}
 \sigma^{\ell}_{ cap}(E_{\rm c.m.};\beta_P, \alpha_T)=
 \sigma^{\ell}_{ qfiss}(E_{\rm c.m.};\beta_P, \alpha_T)
 +\sigma^{\ell}_{ fus}(E_{\rm c.m.}; \beta_P, \alpha_T)
 + \sigma^{\ell}_{ fastfis}(E_{\rm c.m.}; \beta_P, \alpha_T) \label{capt2}
 \end{equation}
where
\begin{equation}
\sigma^{\ell}_{\rm fus}(E_{\rm c.m.})=
\sigma^{\ell}_{cap}(E_{\rm c.m.}) P_{\rm CN}(E_{\rm c.m.},\ell),\label{s_fus}
\end{equation}
is the partial fusion cross section and $P_{CN}(E_{\rm c.m.},\ell)$ is fusion probability as a function of $\ell$. It is clear that the fusion cross section in formula (\ref{s_fus})
includes the cross sections of evaporation residue and fusion-fission products. The fusion cross section, for each orientation angles of the symmetry axes  of the deformed reacting nuclei, is obtained by formula
 \begin{equation}
 \label{totfus}
 \sigma_{fus}(E_{\rm c.m.};\beta_P, \alpha_T)=\sum_{\ell=0}^{\ell_f}(2\ell+1)
 \sigma_{cap}(E_{\rm c.m.},\ell;\beta_P, \alpha_T)
 P_{CN}(E_{\rm c.m.},\ell; \beta_P, \alpha_T).
 \end{equation}
Therefore, taking into account the contributions of all configurations with the orientation angles of the symmetry axes  of the deformed reacting nuclei we
can calculate the averaged value of the fusion cross section  by formula:
\begin{equation}\label{sigmal}
<\sigma_{fus}>_{\{\alpha_{P},\alpha_{T}\}}(E_{c.m.})=\int_0^{\pi/2}\sin\alpha_P\int_0^{\pi/2}
\sin\alpha_T \times \sigma_{fus}(E_{c.m.}; \alpha_P,\alpha_T) d\alpha_Td\alpha_P,
\end{equation}
when both nuclei are deformed nuclei, or by formula:
\begin{equation}\label{sigma2}
<\sigma_{fus}>_{\{\alpha_{T}\}}(E_{c.m.})=\int_0^{\pi/2}
\sin\alpha_T \times \sigma_{fus}(E_{c.m.}; \alpha_T) d\alpha_T,
\end{equation}
for the spherical projectile and deformed target.

Obviously, the quasifission cross section is obtained as
\begin{equation}
 \label{totqfis}
 \sigma_{qfis}(E_{\rm c.m.}; \beta_P, \alpha_T)=
 \sum_{\ell=0}^{\ell_d}(2\ell+1)\sigma_{cap}(E_{\rm c.m.},\ell; \beta_P, \alpha_T)
 (1-P_{CN}(E_{\rm c.m.},\ell; \beta_P, \alpha_T)),
 \end{equation}
where $\ell_d$ is the maximum value of $\ell$ at which the capture events occur.

Another binary process which leads to the formation of two fragments
similar to the ones of fusion-fission or quasifission is the fast fission.
The fast fission cross section  is calculated
by summing the  contributions of the partial cross sections related to the
range $\ell_f\le\ell\le\ell_d$ (for which $B_{\rm f}$=0) leading to the formation of the mononucleus:
 \begin{equation}
 \label{fasfiss}
 \sigma_{fastfis}(E_{\rm c.m.};\beta_P,\alpha_T)=\sum_{\ell_f}^{\ell_d}(2\ell+1)
 \sigma_{cap}(E_{\rm c.m.},\ell; \beta_P, \alpha_T)
 P_{CN}(E_{\rm c.m.},\ell; \beta_P, \alpha_T).
 \end{equation}

The fission and the evaporation residue cross sections are calculated by the advanced statistical
code \cite{epja192004,epja222004,mpla202005} that takes into account the damping
of the shell correction in the fission barrier as a function of
nuclear temperature and orbital angular momentum.

\subsection{Calculation of competition between quasifission and complete fusion processes}

The fusion cross section is calculated from the branching ratio $P_{\rm CN}(Z)$ of the decay rates of overflowing the border of the potential well ($B^{(Z)}_{\rm qf}$) along $R$ at
a given mass asymmetry (decay of DNS in quasifission fragments) over the barriers on mass asymmetry axis $B^{*}_{\rm fus}$ for the complete fusion or $B^{(Z)}_{\rm sym}$ in opposite
direction to the symmetric configuration of DNS (for details see Fig. 4 of Ref. \cite{prc812010} ):
\begin{equation}
\label{PcnG}
P^{(Z)}_{\rm CN}(E^*_{DNS})\approx{\frac{{\Gamma^{(Z)}_{\rm fus}}(B^{*}_{\rm fus},E^*_{DNS})}{{\Gamma^{(Z)}_{(\rm qf)}(B_{\rm qf},E^*_{DNS})+\Gamma^{(Z)}_{(\rm fus)}(B^{*}_{\rm fus},E^*_{DNS})+\Gamma^{(Z)}_{\rm sym}(B_{\rm sym},E^*_{DNS})}}},
\end{equation}
where $\Gamma_{\rm fus}$,  $\Gamma_{\rm qf}$ and  $\Gamma_{\rm sym}$ are corresponding widths determined by the level densities on the barriers  $B^{*}_{\rm fus}$, $B^{*}_{\rm sym}$
and $B_{\rm qf}$ involved in the calculation of $P_{\rm CN}$ are used in the model \cite{prc722005,NasirovNPA759, AdamianPRC2003} based on the dinuclear system concept \cite{Volkov}.
 Here $E^{*}_{\rm DNS}(Z_P,A_P,\ell)= E_{\rm c.m.}-
V(Z_P,A_P,\ell,R_{\rm m})$ is the excitation energy of dinuclear system in the entrance channel, where $Z_P$ and $A_P$ are charge and mass numbers of the projectile nucleus. $V(Z,A,R_{\rm m},\ell)$
is the minimum value of the nucleus-nucleus potential well (for the DNS with charge asymmetry $Z$) and  its position on the relative distance between the centers of nuclei is marked as $R=R_{\rm m}$.
The value of $B_{\rm qf}$ for the decay of DNS with the given charge asymmetry of fragments is equal to the depth of the potential well in the nuclear-nuclear  interaction. The intrinsic fusion barrier
$B^{*}_{\rm fus}$ is connected with mass (charge) asymmetry degree of freedom of the dinuclear system and it is determined from the potential energy surface:
\begin{equation}
U(Z;R,\ell)=U(Z,\ell,\beta_{1},\alpha_{1};\beta_{2},\alpha_{2}) =B_{1}
+B_{2}+V(Z,\ell,\beta_{1},\alpha_{1};\beta_{2},\alpha_{2};R)-
(B_{CN}+V_{CN}(\ell)).\label{Udz}
\end{equation}
Here, $B_{1}$, $B_{2}$ and $B_{\rm CN}$ are the binding energies of the nuclei in DNS and the CN, respectively, which were obtained from \cite{MollerADND1988}; the fragment deformation parameters
$\beta _{i}$ are taken from the tables in \cite{RamanADND1987, SpearADND1989, MollerADND1988} and $%
\alpha _{i}$ are the orientation angles of the reacting nuclei
relative to the beam direction; $V_{\rm CN}(\ell)$ is the
rotational energy of the CN. The distribution of neutrons between
two fragments for the given proton numbers $Z$ and $Z_{2}$ or
ratios $A/Z$ and $A_{2}/Z_{2}$ for both fragments were determined by minimizing the potential $U(Z;R)$ as a function of $A$ for each $Z$.

The driving potential $U_{\rm dr}(Z)\equiv U(Z,R_{m})$ is a curve
linking minima in the potential well corresponding to each charge
asymmetry $Z$ in the valley of the potential energy surface from
$Z=0$ up to $Z=Z_{\rm CN}$ as a function of the relative distance
(see  Fig. 4 of Ref. \cite{prc812010}).  We define the intrinsic
fusion barrier for the dinuclear system with charge asymmetry $Z$
as $B_{\rm fus}^{\ast }(Z,\ell)= U(Z_{\rm max},R_{\rm m}(Z_{\rm
max}),\ell)-U(Z,R_{\rm m}(Z),\ell)$, where $U(Z_{\rm max},\ell)$
is a maximum value of potential energy at  $Z=Z_{\rm max}$ in the
valley along the way of complete fusion from the given $Z$
configuration. The $B_{\rm sym}^{\ast }(Z,\ell)$ is defined by the
similar way as shown in Fig. 4 c of Ref. \cite{prc812010}.

The masses and charges of the projectile and target nuclei are dynamical variables during capture and after formation of the DNS. The intense proton and neutron exchange between constituents of DNS is taken into account by
calculation of the complete fusion probability $P_{\rm CN}$ as fusion from all
populated DNS configurations according to the formula
\begin{equation}
\label{PcnY}
P_{CN}(E^{*}_{\rm DNS}(Z,A,\ell);\{\alpha_{i}\})=\sum\limits_{Z_{\rm sym}}^{Z_{\rm max}}Y_Z(E^{*}_{\rm DNS}(Z,A,\ell))P^{(Z)}_{\rm CN}(E^{*}_{\rm DNS}(Z,A,\ell);\{\alpha_{i}\})
\end{equation}
where $E^{*}_{\rm DNS}(Z,A,\ell)= E^{*}_{\rm DNS}(Z_P,A_P,\ell)+\Delta{Q_{\rm gg}(Z)}$ is the excitation energy of DNS with angular momentum $\ell$ for a given value of its charge-asymmetry configuration $Z$ and $Z_{\rm CN}-Z$;
$Z_{\rm sym}=(Z_1+Z_2)/2$;  $\Delta{Q_{\rm gg}}(Z)$ is the change of $Q_{\rm gg}$-value by changing the charge (mass) asymmetry of DNS; $Y_Z(E^{*(Z)}_{\rm DNS})$ is the probability of
population of the ($Z, Z_{\rm CN}$-Z) configuration  at $E^{*(Z)}_{\rm DNS}$ and given orientation angles ($\alpha_1,\alpha_2$). $Y_Z(E^*_{DNS},\ell,t)$ is the probability of
population of the configuration $(Z, Z_{tot}-Z)$ at $E^*_{DNS}(Z)$
and $\ell$. The evolution of $Y_Z$  is calculated by solving the
transport master equation:
\begin{eqnarray}
\label{massdec}
\frac{\partial}{dt}Y_{Z}(E^*_Z,\ell,t)&=&\Delta^{(-)}_{Z+1}
Y_{Z+1}(E^*_Z,\ell,t)+
\Delta^{(+)}_{Z-1}  Y_{Z-1}(E^*_Z,\ell,t)\nonumber\\
&&-(\Delta^{(-)}_{Z}+\Delta^{(+)}_{Z}+\Lambda^{qf}_{Z})
Y_{Z}(E^*_Z,\ell,t), \hspace*{0.1cm}\mbox{\rm for} \ Z=2,3,...,
Z_{tot}-2.
\end{eqnarray}
Here, the transition coefficients of multinucleon transfer  are
calculated as in \cite{Jolos86}
\begin{eqnarray}
\label{delt} \Delta^{(\pm)}_{Z}=\frac{1}{\Delta t}
\sum\limits_{P,T}|g^{(Z)}_{PT}|^2 \ n^{(Z)}_{T,P}(t) \
(1-n^{(Z)}_{P,T}(t)) \
 \frac{\sin^2(\Delta t(\tilde\varepsilon_{P_Z}-
 \tilde\varepsilon_{T_Z})/2\hbar)}{(\tilde\varepsilon_{P_Z}-
 \tilde\varepsilon_{T_Z})^2/4},
\end{eqnarray}
where the matrix elements $g_{PT}$  describe one-nucleon exchange
between the nuclei of DNS, and their values are calculated
microscopically using the expression obtained  in
Ref.~\cite{AdaS92}. A non-equilibrium distribution of the
excitation energy between the fragments was taken  into account in
calculations of the single-particle occupation numbers $n_{P}$ and
$n_{T}$ as it was done in Ref.~\cite{Adam94};
 $\tilde\varepsilon_{P_Z}$ and $\tilde\varepsilon_{T_Z}$
are perturbed energies of single-particle states. In Eq.
\ref{massdec}, $\Lambda^{qf}_{Z}$ is the Kramers rate for the
decay probability of the dinuclear system into two fragments with
charge numbers
 $Z$ and $Z_{tot}-Z$   (details in Ref. \cite{Adam03}), and it is
 proportional to  $\exp\left(-B_{qf}(Z)/(kT)\right)$
where $B_{qf}(Z)$ is the quasifission barrier.
\begin{figure}[htb]
\vspace*{2.5cm}
\begin{flushleft}
\resizebox{1.1\textwidth}{!}{\includegraphics{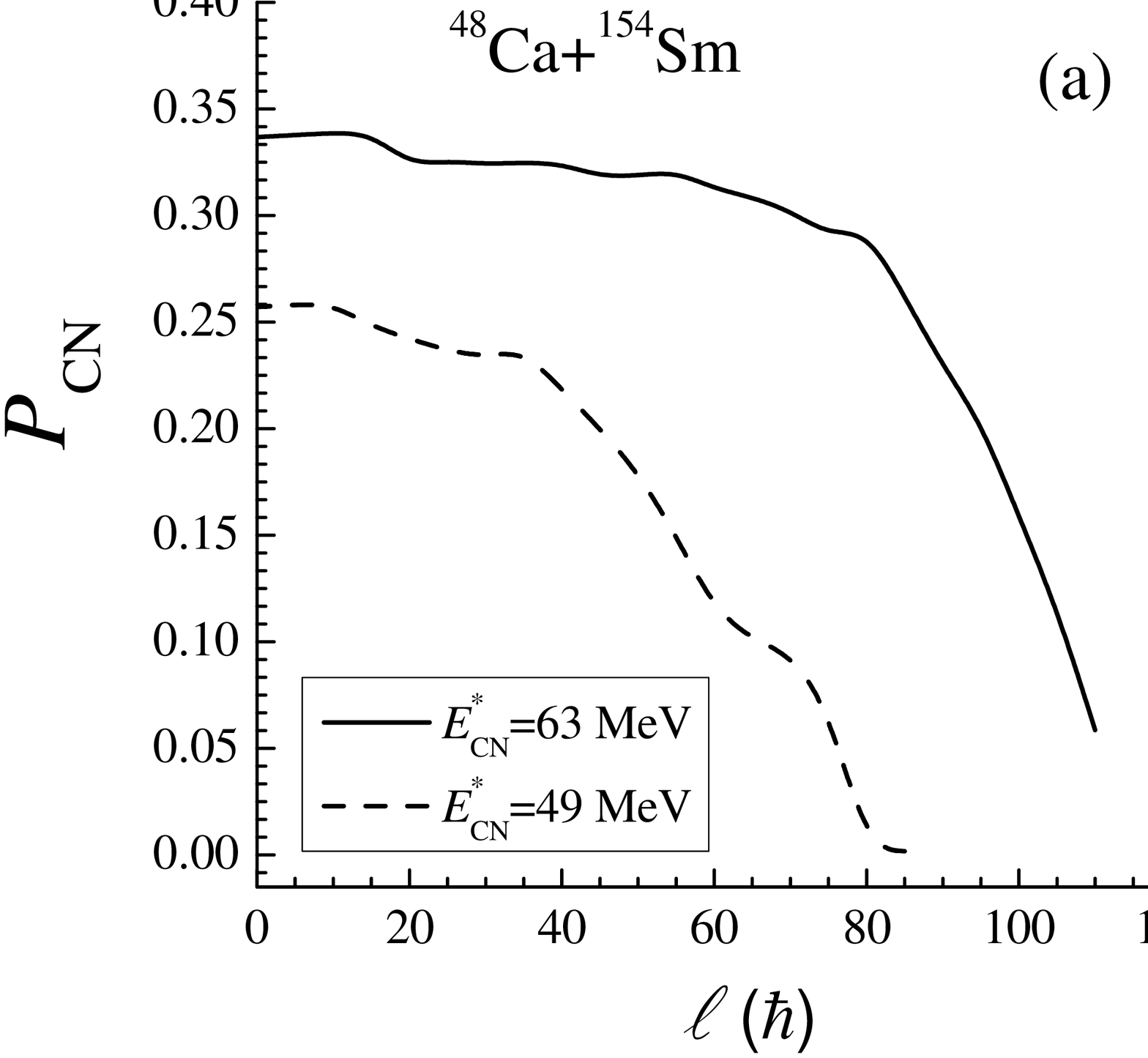}\includegraphics{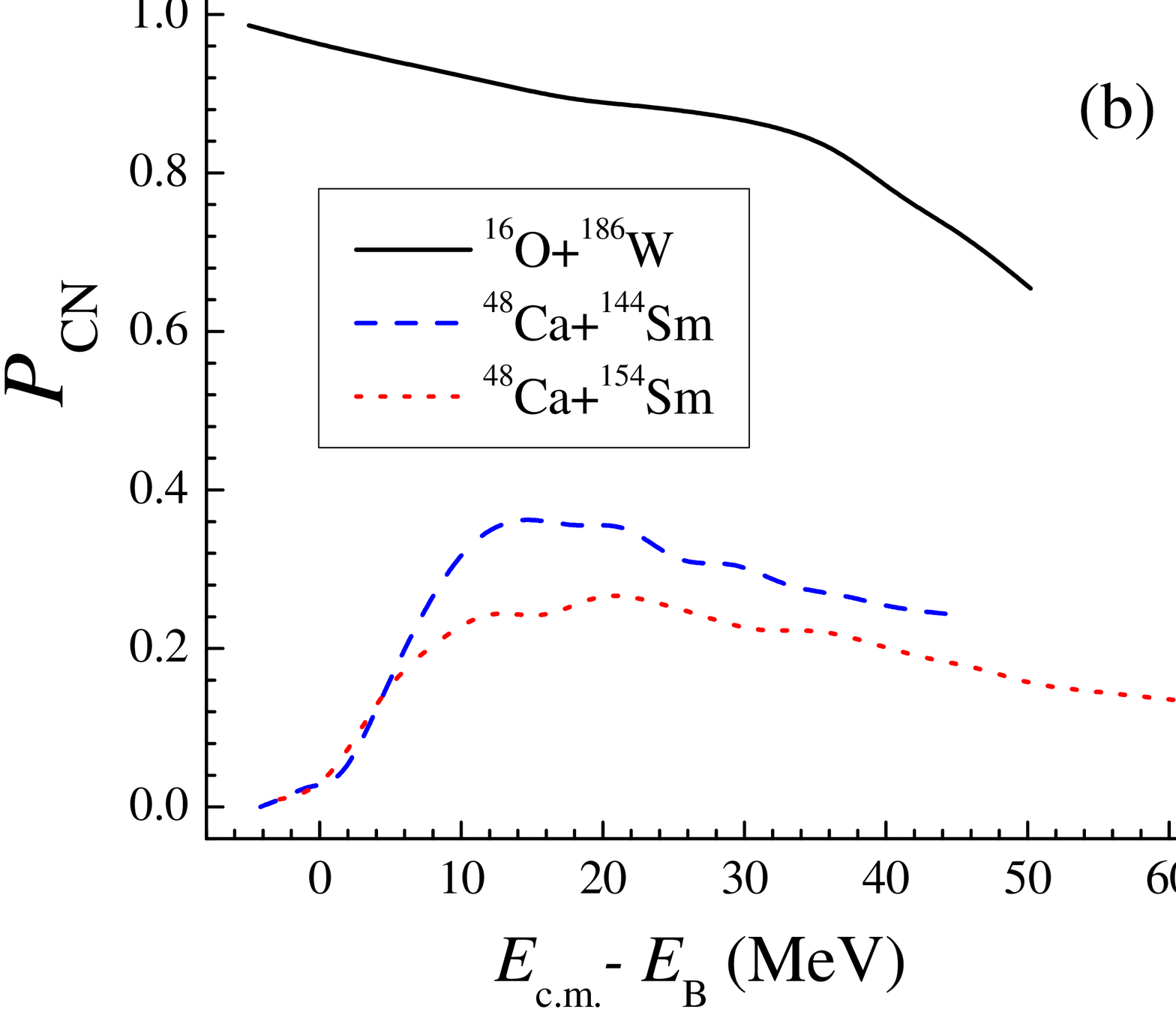}}% Here is how to import EPS art
\end{flushleft}
\vspace*{-3.3cm} \caption{\label{PcnFig} (a) The probability $P_{\rm CN}$
of compound nucleus formation  as a function of the angular
momentum
 of dinuclear system $\ell$ at energies $E_{\rm c.m.}$ =138
and 154 MeV, corresponding to the excitation energies of the
compound nucleus $E^*_{\rm CN}$=49 and 63 MeV. (b) The DNS model results
for the fusion probability  $P_{CN}$ for the $^{16}$O+$^{186}$W,
$^{48}$Ca+$^{144}$Sm  and $^{48}$Ca+$^{154}$Sm reactions as a
function of the collision energy relative to the interaction
barriers corresponding to each of reactions.}
\end{figure}

\begin{SCfigure}[][tb]
%\vspace*{3.0cm}
%\par
\centering
\includegraphics[width=0.7\textwidth]{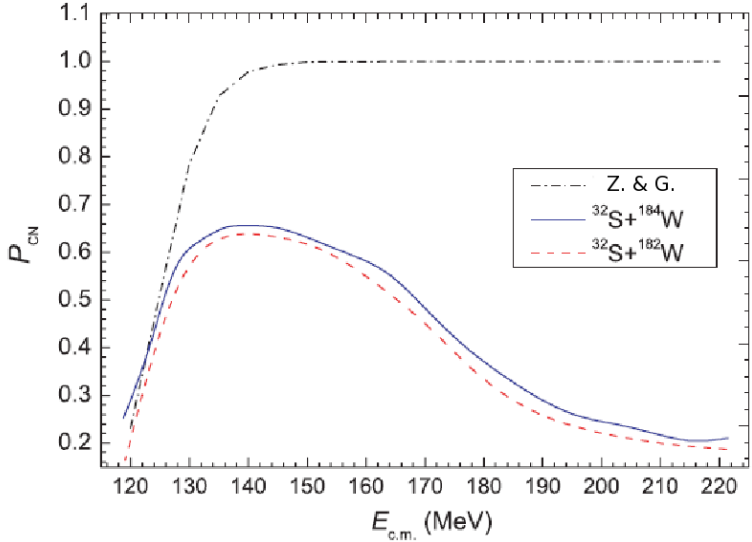}
%Fig6
% Here is how to import EPS art
\vspace{-1.35cm}
\caption{Our theoretical values of the fusion
probability $P_{\rm CN}$ for the $^{32}$S+$^{184}$W reaction as a function of
collision energy $E_{\rm c.m.}$, and the one of other authors (Z.~\&~G.) of Ref. \cite{prc782008}. }\vspace{0.35cm}
\label{PcnEcm}
\end{SCfigure}

Eqs. (\ref{massdec})  with the coefficients (\ref{delt}) and
initial condition $Y_{Z}(E^*,0)=\delta_{Z,Z_P}$ are solved
numerically and  the primary mass and charge  distributions are
found for a given interaction time $t_{int}=5\cdot10^{-21}$s (see
Ref.~\cite{Toke85}). In Eq. (\ref{PcnY}), we use the
definition $Y_{Z}(E^*_Z,\ell)=Y_{Z}(E^*_Z,\ell,t_{int}(\ell))$. In
(\ref{delt}) we use $\Delta t=10^{-22}$s$\,<< \,t_{int}$.

As an example of our calculation of the fusion probability $P_{\rm CN}$ versus the angular momentum $\ell$, we present the results for the $^{48}$Ca+$^{154}$Sm reaction at two
different $E^*_{\rm CN}$ values (see Fig. \ref{PcnFig} (a))
and the function $P_{\rm CN}$ versus the energy $E_{\rm c.m.}-E_{\rm B}$ for the reaction that lead to different isotopes of thorium (see Fig. \ref{PcnFig} (b)).
This last figure shows the effect of the entrance channel on the $P_{\rm CN}$ fusion probability.

Moreover, Fig. \ref{PcnEcm} shows the calculation of $P_{\rm CN}$ as a function of the energy $E_{\rm c.m.}$ for the two close $^{32}$S+$^{182}$W and $^{32}$S+$^{184}$W
reactions for a wide range of excitation energies, in comparison with the calculation reported in Ref. \cite{prc782008}. For these two close reactions we observe two region
of strong hindrance to fusion at very small and large values of the collision energy of $E_{\rm c.m.}$ for both reactions while the approximate formula
used in Ref. \cite{prc782008} gives similar results only at small energies, but  $P_{\rm CN}$ increase quickly reaching a maximum value equal to 1 at $E_{\rm c.m.}$ = 145 MeV.
In addition, there is no difference between the values of $P_{\rm CN}$ calculated by the approximated formula for the $^{32}$S+$^{182}$W and $^{32}$S+$^{184}$W reactions.

\subsection{Compound Nucleus and Fast Fission contributions from the complete fusion formation}

The branching ratio of the quasifission, fusion, and fast fission cross sections in formula  (\ref{capt2})  depend on the masses and charges, shell structure, mass asymmetry
of projectile-target nuclei, as well as on the energy and impact parameter of heavy ion collision.  As it was noted in Section \ref{intro}, the formation
of quasifission and fast fission products bypasses the stage of
the CN formation. The difference between the two mentioned processes is in
their dependence on the orbital angular momentum: quasifission is
possible at all angular momenta of collision as fast fission
occurs only at $\ell>\ell_f$, where $\ell_f$ is the angular
momentum value at which the fission barrier of the compound
nucleus disappears.
\begin{SCfigure}[][tb]
\label{Fig1} \vspace{-3.0cm}
\centering
\includegraphics[width=0.7\textwidth]{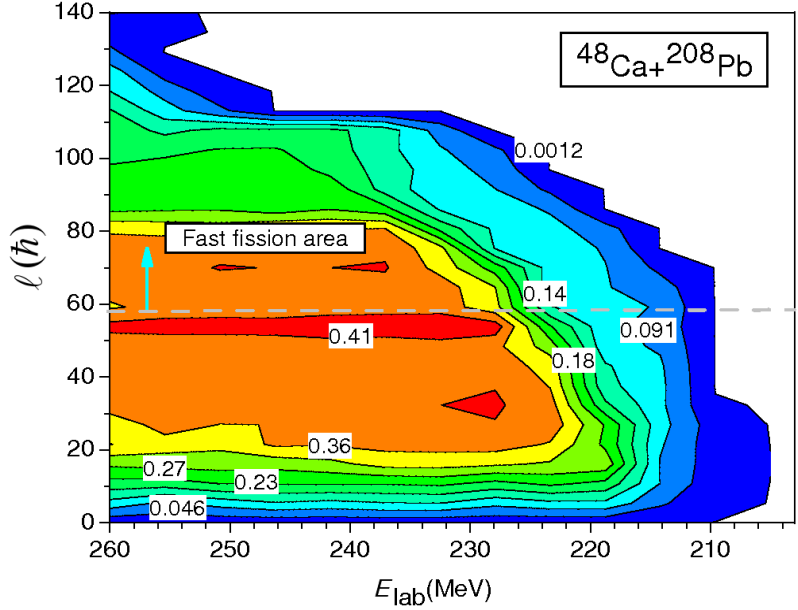} \vspace*{-0.3cm}
\caption{The results of calculation for the fusion angular
momentum distribution of  the $^{48}$Ca+$^{208}$Pb reaction. The
upper part of the figure, labelled by "Fast fission area",
indicates the region where also the fast fission process appear. \label{fasfis}}\vspace{2.5cm}
\end{SCfigure}
The mass distributions of
products of the capture reactions can be mixed by different
proportions in different mass regions.
To avoid ambiguity, the
authors of Ref. \cite{Bock} referred to the registered
products as ones of the symmetric fragmentation (capture) because
the genuine fission process is the decay of the compound nucleus
into two fragments.
 The mass distribution of products of the
fast fission reactions can be alike to that of the fusion-fission
reactions. But the angular distribution of products of the former
reaction is more anisotropic than that of the latter reaction due
to the difference in their reaction time. In order to estimate the
contribution of the fast fission into the calculated capture cross
section, we use the value $\ell_f=58$ obtained in Ref.
\cite{Bock} for the $^{48}$Ca+$^{208}$Pb reaction.

In Fig. \ref{fasfis}
the results of calculation for the fusion angular momentum
distribution of this reaction is presented. One can see that
maximum of the partial fusion cross section is around $\ell=50$.
The fast fission contribution for $\ell > \ell_{\rm f}$ is calculated by formula (\ref{Udz}), while the fusion cross section of CN must include  the evaporation residues and fusion-fission cross sections only.
In our calculation of evaporation residues, the
values of the angular momentum $\ell < 58$ were used.

\subsection{Evaporation Residue}

The probability of the
formation of a evaporation residue nucleus surviving fission
with mass number  $A=A_{CN}-(\nu(x)+y(x)+4k(x))$ and charge
number $Z=Z_{CN}-(y(x)+2k(x))$ from a heated and rotated compound
nucleus $^{A_{CN}}Z_{_{CN}}$ after emissions of $\nu$ neutrons,
$y$ protons, and $k~\alpha$-particles at the $x$th  step of
the de-excitation cascade is determined by the formula \cite{epja222004,mpla202005}:

\begin{equation}
\label{evapor}
\sigma_{ER(x)}(E^*_x)=\sum_{\ell=0}^{\ell_f}(2\ell+1)
\sigma^{\ell}_{(x-1)}(E^*_x)W_{{\rm sur}(x-1)}(E^*_x,\ell),
\end{equation}
where $\sigma^{\ell}_{(x-1)}(E^*_x)$ is the partial  cross section
of the  intermediate nucleus formation at the $(x-1)$th step and
$W_{{\rm sur}(x-1)}(E^*_x,\ell)$ is the survival probability of
the $(x-1)$th intermediate nucleus against fission along the
de-excitation cascade of CN;
$\ell_f$ is the value of angular momentum $\ell$ at which
the fission barrier for a compound nucleus
disappears completely \cite{Sierk86};
$E^*_x$ is the  excitation energy of
the nucleus formed at the $x$th step of the de-excitation cascade.
 It is clear that  $\sigma^{\ell}_{(0)}(E^*_0)=\sigma^{\ell}_{\rm fus}(E^*_{CN})$ at
\begin{equation}
\label{Ecn}
E^*_{CN}=E^*_0=E_{\rm c.m.}+Q_{gg}-E_{rot},
\end{equation}
 where $E_{\rm c.m.}$, $Q_{gg}$,
and $E_{rot}$ are the collision energy in the center of mass system,
the reaction $Q_{gg}$-value, and rotational energy of the compound nucleus, respectively.
 The numbers of the emitted neutrons, protons, $\alpha$-particles and
$\gamma$-quanta, $\nu(x)$n, $y(x)$p, $k(x)\alpha$, and
$s(x)\gamma$, respectively, are functions of the step $x$. The
emission branching ratios of these particles depend on the
excitation energy $E^*_{A}$ and angular momentum $\ell_{A}$ of the cooling
intermediate nucleus.
\begin{figure}[th]
\begin{center}
%\vspace{-1.2cm}
\hspace{-10cm}\resizebox{0.40\textwidth}{!}{\includegraphics{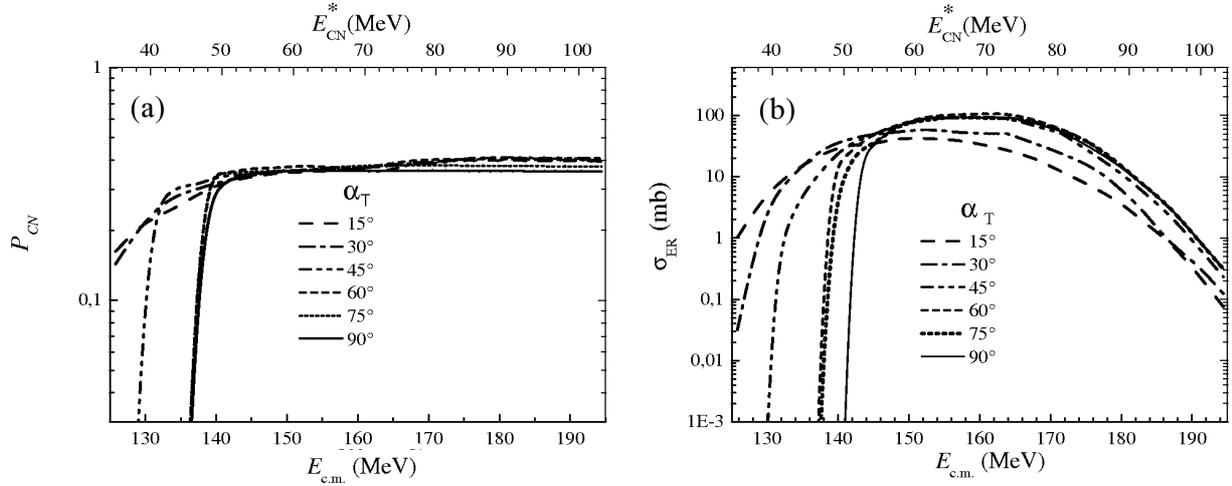}}
\end{center}
%\vspace{-4.8cm}
\vspace{-0.3cm}
\caption{Fusion probability $P_{CN}$ (a) versus the collision energy
$E_{\rm c.m.}$ for different values of the $\alpha_T$ angle.  Evaporation residue cross sections (b) versus the
$E_{\rm c.m.}$ energy for various orientation angles $\alpha_T$
 of the $^{154}$Sm target.}
\vspace{-0.1cm}
\label{pcnorien}
\end{figure}

The de-excitation cascade, characterized by the emission of the
above-mentioned particles, starts from the compound nucleus $^{A_{CN}}Z_{_{CN}}$.
 Its formation probability
is the partial cross section of complete fusion
$\sigma^{\ell}_{\rm fus}(E^*_{CN})$ corresponding to the orbital
angular momentum $\ell$. The fusion cross section
is equal to the capture cross section for light systems or light projectile-induced  reactions, while for  reactions
with massive nuclei, it becomes a model-dependent   quantity.
Concerning estimation
of the fusion cross section from the experimental fragments data,
its value is sometimes an ambiguous
quantity because of difficulties in separating  the fusion-fission fragments from
the quasifission fragments in the case of overlap of mass and angular distributions.

As an example we present the results of our calculation for the $^{48}$Ca+$^{154}$Sm reaction regarding the fusion probability $P_{\rm\, CN}$ (see Fig. \ref{pcnorien}
(a)) and evaporation residue cross sections $\sigma_{\rm\, ER}$ (see Fig. \ref{pcnorien}~\,(b)) as a function of the collision energy  $E_{\rm\, c.m.}$ for different values of the target orientation angles $\alpha_{\rm T}$.
 The dependence of the quasifission-fusion competition during the
evolution of the dinuclear system and the sensitivity of the
fission-evaporation competition during the de-excitation cascade of the
compound nucleus on the values of orientation angle  $\alpha_T$
are demonstrated. The analysis of the dependence of the compound nucleus and
evaporation residue formation cross sections on  $\alpha_T$  shows that
the observed yield of evaporation residues in the $^{48}$Ca+$^{154}$Sm reaction
at the low energies ($E_{\rm c.m.}<$137 MeV) is formed in the collisions
$\alpha_T<45^{\circ}$. Because the initial beam energy is enough
to overcome the corresponding Coulomb barrier for the collisions with these
orientation angles $\alpha_T$. Only in this case it is
possible  formation of dinuclear system
which evolves to compound nucleus or breaks up into two fragments after multinucleon
exchange without formation of the compound nucleus.
  At larger energies (about $E_{\rm c.m.}=$140--180 MeV) all
 orientation angles of the target-nucleus can contribute
to $\sigma_{ER}$ and its values are in the 10--100 mb range.
At the larger collision energies ($E_{c.m.}>$158
MeV) the complete fusion still increases
 but the evaporation residue cross section $\sigma_{ER}$ goes
 down and its values are in the 1--0.1 mb range due to the
 strong decrease of the survival probability of the heated
 compound nucleus along de-excitation cascade. This is connected
 by the decrease of the  fission barrier for a compound nucleus
 by increasing its
excitation energy \cite{epja192004,jpsj72} and angular momentum \cite{Sierk86}.

Another phenomenon leading to decrease of $\sigma_{ER}$ at  higher
 energy is the fast fission process which is the splitting of the mononucleus
into two fragments due to absence of the fission barrier  at very high the angular
momentum  $\ell > \ell_f$.

\section{Mass distribution for quasifission and fusion-fission fragments at different energies}

\begin{SCfigure}[][htb]
% Fig.10
\vspace{9.5cm}
\centering
\includegraphics[width=0.7\textwidth]{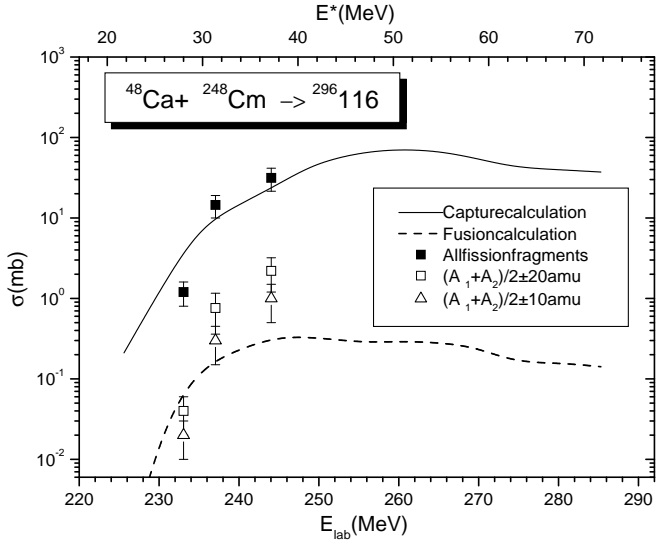}
\vspace{-13.3cm} \caption{\label{116cf}The capture and fusion
cross section calculation, in comparison with the experimental
data \cite{Itkis01,Kozulin02}, for the $^{48}$Ca + $^{248}$Cm
reaction leading to the $^{296}$116 superheavy CN. The difference
between the fusion cross section and more symmetric fragment yield
(($A_1+A_2)/2\pm 10$ amu) at a higher excitation energy is related
to the contribution of the quasifission process yielding more
symmetric fragments.} \vspace*{10pt}
\end{SCfigure}

To show the importance of the overlap of mass distribution of quasifission and fusion-fission products in the unambiguous sorting out  of events belonging to true
fusion-fission process from the ones of quasifission process, we compare
theoretical and experimental results of capture and fusion cross section
for the $^{48}$Ca+$^{248}$Cm reaction in Fig. \ref{116cf} and discuss these results.
%\vspace*{-0.4cm}

The calculated capture cross section (solid line in Fig.\ref{116cf}) is in good
agreement with the experimental data \cite{Itkis01}  for the
production of all fragments (full squares), while the theoretical fusion cross
section (dashed line) is not in agreement with the experimental data obtained assuming the masses of the fusion-fission fragments belong in the ranges $(A_1+A_2)/2\pm 10$ or ($(A_1+A_2)/2\pm 20$.

The presence of the quasifission contribution into measured fusion-fission cross section in the ranges under discussion causes disagreement with the theoretical results.
If it is assumed that the experimental fusion - fission events are in the $(A_1+A_2)/2\pm 10$ amu interval (almost close to the
$\sqrt{(A_1+A_2)/2}$ value), the calculated fusion cross sections
(the dashed line) will be closer to  the new set of the
experimental data (open triangles \cite{Kozulin02} in
Fig.\ref{116cf}). Indeed, in this case there is an appreciable
contribution of the quasifission process (or a contribution which
cannot be neglected), in addition to the fusion-fission fragment
formation. Therefore, the estimated experimental fusion cross
section, connected with the new set of the experimental events of
fission fragments, still appears to be a little larger than the
calculated fusion excitation function at higher excitation
energies. A preliminary calculation of the mass distribution of
quasifission fragments for a fixed reaction time $t_{reac}$ of a
DNS performed in the framework of the model developed on the basis
of the dinuclear system concept \cite{Jolos89,nasmess}, indicates
that the fragments of the quasifission process also appear in the
mass-symmetric region and are mixed with the fragments coming from
the fusion-fission process.

\section{Study of the reaction dynamics of two close mass asymmetric reactions by the analysis of the fusion-evaporation and fusion-fission processes}

 \begin{figure}[h]
% Fig.3
\vspace*{3.85cm}
\begin{center}
\resizebox{0.75\textwidth}{!}{\includegraphics{{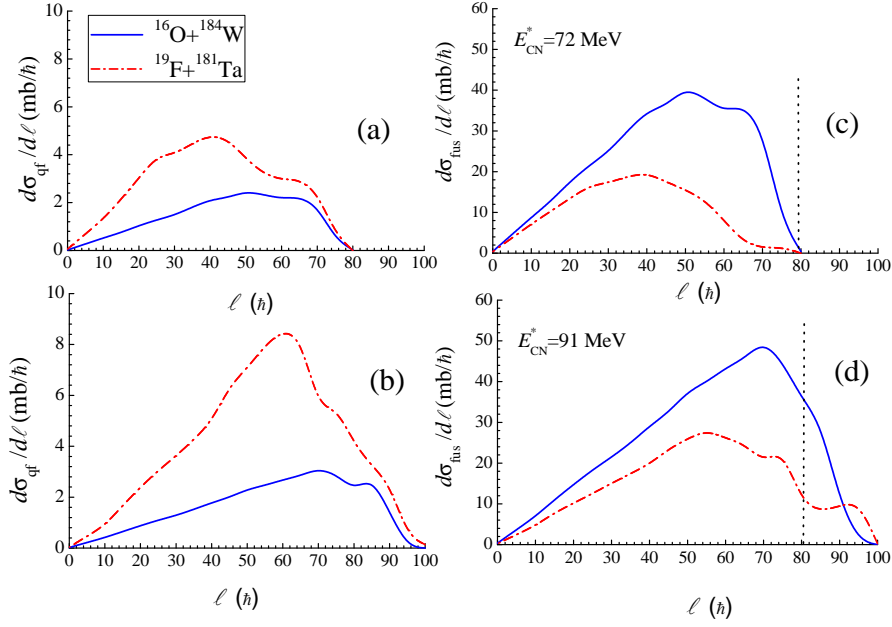}}}% Here is how to import EPS art
\vspace*{-3.7 cm} \caption{\label{fig3} Partial quasifission ((a) and (b) panels) and fusion ((c) and (d) panels) cross sections as a function of the angular momentum $\ell$
for the $^{16}$O+$^{184}$W (solid line) and $^{19}$F+$^{181}$Ta
(dashed line) reactions.  In the panel (a) the beam energies of reactions induced by $^{16}$O and  $^{19}$F are $E_{\rm c.m.}=$96.2 MeV and 95.7 MeV, respectively.
These energies correspond to the same excitation energy of CN $E^*_{\rm CN}=$72 MeV for the both systems shown in the panel (c).
Analogously, in the panel (b) the beam energies are $E_{\rm c.m.}=$115.2 MeV  and  $E_{\rm c.m.}=$114.7 MeV, respectively. The corresponding  excitation energy of CN is  91 MeV shown in the panel (d).
In (c)and (d), the vertical dashed line at $\ell_{\rm f}=80 \hbar$ separates
compound nucleus and fast fission contributions.
}\vspace*{-0.5 cm}
\end{center}
\end{figure}

There are three main processes causing  hindrances to  ER formation
in reactions with massive nuclei:  quasifission, fusion-fission, and  fast fission  \cite{prc792009n}.  All of these processes produce binary fragments in different stages of reaction.
Moreover, the  angular and mass distributions of some parts of their products can overlap \cite{prc792009n,anisEPJA34}. Ignoring this mixing may lead to ambiguity at analysis of the
experimental data connected with the binary fragments. This problem should be studied carefully.

The ER formation process is often considered as the third stage of the three-stage process. The first stage is  capture--formation of the  DNS  after full momentum transfer into
the deformation energy of nuclei, their excitation energy, and rotational energy from the initial relative motion of the colliding heavy ions in the center-of-mass system. Capture
 takes place if the initial energy of the projectile in the center-of-mass system is enough to overcome the interaction
barrier (Coulomb barrier + rotational energy of the entrance channel) \cite{NasirovNPA759}.
The study of the dynamics of  heavy ion collisions at energies near
the Coulomb barrier  shows that complete fusion does not occur immediately in collisions of massive nuclei \cite{epja192004,Back31,prc792009n,VolPLB1995,AdamianPRC2003}.
After formation of the  DNS, the quasifission process competes with the formation of CN. Quasifission occurs when the  DNS prefers to break down into fragments instead of
being transformed into a fully equilibrated  CN. The number of events  contributing to quasifission increases drastically by increasing  the sum of the Coulomb interaction
 and rotational energy in the entrance channel \cite{epja82000,prc722005,prc792009}.

Another reason for the  decreasing yield of ER with increasing excitation energies  is the usual fission of a heated and rotating CN  that was formed in competition with quasifission. The stability of a massive CN
decreases due to the decrease  in the fission barrier by increasing
its excitation energy $E^*_{CN}$ and angular momentum $\ell$
\cite{ArrigoPRC1992,ArrigoPRC1994,SagJPG1998}.
%\begin{SCfigure}[][tb]
\begin{figure}[tb]
% Fig.4
\vspace*{-2.64cm}
\centering
\includegraphics[width=0.8\textwidth]{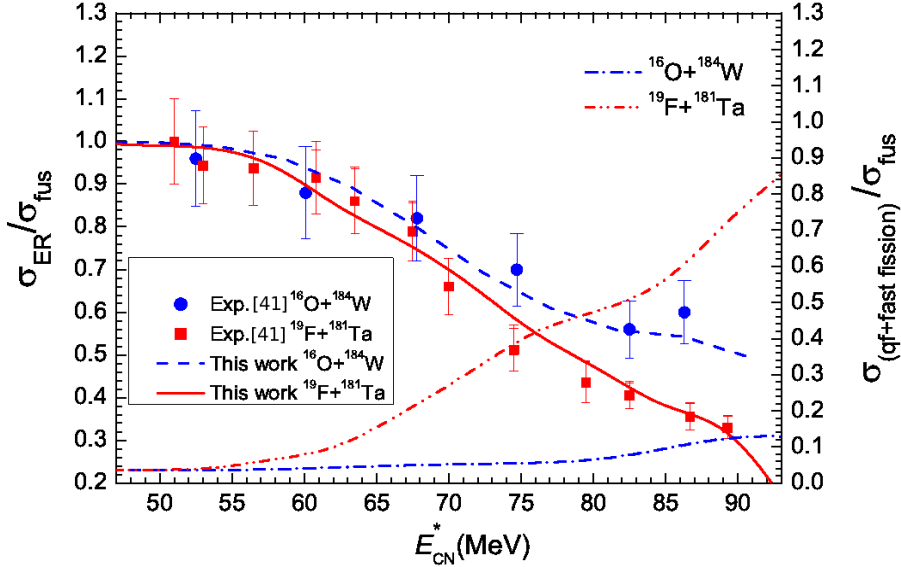}% Here is how to import EPS art
\vspace*{-0.75 cm} \caption{\label{fig4} Comparison of the
experimental values of the evaporation residue cross sections
normalized with respect to the capture cross sections for the
$^{16}$O+$^{184}$W (solid circles) \cite{ShidlingPLB}
and $^{19}$F+$^{181}$Ta systems
(solid squares) \cite{ShidlingPLB} with the corresponding
theoretical results (dashed and solid lines, respectively) as a
function of the excitation energy $E^*_{\rm CN}$ of CN
(left axis). Theoretical results of the sum of the quasifission and
fast fission cross sections (normalized with respect of the fusion
cross sections) for the $^{16}$O+$^{184}$W (dot dashed line) and
$^{19}$F+$^{181}$Ta (dot-dot dashed line) systems are presented
versus $E^*_{\rm CN}$ and compared on the right axis.}% \vspace{3.2cm}
\end{figure}
\begin{figure}[h]
% Fig.5
\vspace*{4.0cm}
\begin{center}
\resizebox{0.85\textwidth}{!}{\includegraphics{{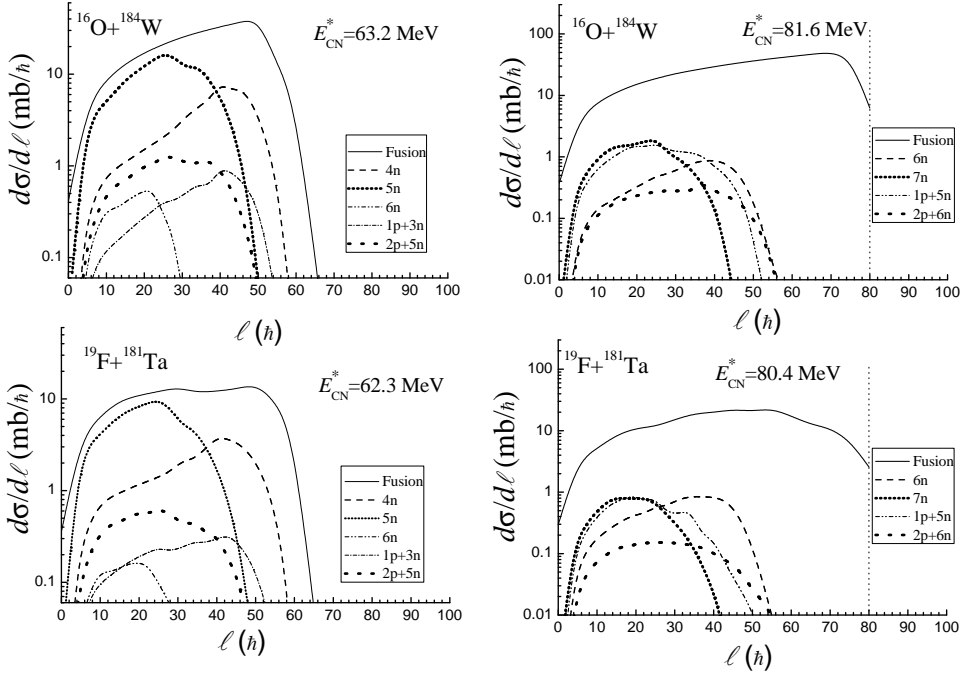}}}% Here is how to import EPS art
\vspace*{-4.25 cm} \caption{\label{fig5} Spin distribution of
evaporation residue cross sections as a function of the angular
momentum $\ell$. The upper part is for  the $^{16}$O+$^{184}$W
reaction, the lower part is for $^{19}$F+$^{181}$Ta reaction, both
at two $E^*_{\rm CN}$ energies:  approximately 62-63 MeV and 80-81 MeV.}
\end{center}
\end{figure}%\vspace*{-3.55 cm}
The theoretical values of the quasifission partial cross sections for the $^{19}$F+$^{181}$Ta and  $^{16}$O+$^{184}$W reactions are presented
in the left panels of Fig. \ref{fig3}. It is seen from these figures that
 the quasifission takes place at all values of $\ell$ leading to capture.
The angular momentum distributions of CN formed in these reactions
at the excitation energies $E^*_{\rm CN}= $72 and 91 MeV are
presented in the right panels of Fig. \ref{fig3}. The spin distributions of CN formed in each of these reactions  differ mainly by the probability but not by the values of
the angular momentum ranges. This means that the number of CN formed in both reactions under discussion are different, but they have a similar range of the angular momentum $\ell$.
The vertical dotted lines at $\ell_{f}=80\hbar$  in these panels separates the complete fusion  ($\ell_{f}<80\hbar$) and fast fission  ($\ell_{f}\ge 80\hbar$) regions of the angular momentum.

The quasifission and fast fission processes produce binary fragments which
can overlap with  those of the fusion-fission channel and the amount of mixed detected fragments  depends on the mass asymmetry of entrance channel, as well as on the shell structure of the
 reaction fragments being formed. The suggestion for the experimental studies of the difference between characteristics of the fusion-fission, quasifission and fast fission
 products can be made when their mass (charge), kinetic energy and angular distributions are explored in detail by dynamical calculations allowing to obtain the relaxation times of these processes.
Therefore,  the correct estimation of the CN formation probability  in the reactions with massive nuclei is a difficult task for both experimentalists and theorists. Different assumptions about the fusion process are used
in different theoretical models which can give different cross sections.
The experimental methods used to estimate the fusion
probability depend on  an unambiguous identification of the
complete fusion products among the quasifission products. The
difficulties arise when the mass (charge) and angular
distributions of the quasifission and fusion-fission fragments
strongly overlap, depending on the reaction dynamics.
As a result, the complete fusion cross sections may be overestimated
\cite{prc792009}.

We confirm that the compared ratios of the cross sections between evaporation residues and complete fusion $\sigma_{ER}/\sigma_{fus}$
for the $^{16}$O+$^{184}$W and $^{19}$F+$^{181}$Ta reactions
discussed in \cite{ShidlingPLB} are not free from the above-mentioned
ambiguity in the determination of the fusion cross section $\sigma_{fus}$.
Theoretical values of the fusion cross section include only evaporation residues and fusion-fission cross sections
\begin{equation}
\label{fusion}
\sigma_{\rm fus}=\sigma_{\rm ER}+\sigma_{\rm ff}.
\end{equation}
The  experimental values of fusion cross section reconstructed from the detected  fissionlike fragments and evaporation residues \cite{ShidlingPLB}:
\begin{equation}
\label{expfus}
\sigma^{\rm(exp)}_{\rm fus}=\sigma_{\rm ff}+
\sigma_{\rm qf}+\sigma_{\rm fast\,fis}+\sigma_{\rm ER},
\end{equation}
where $\sigma_{\rm ff}$, $\sigma_{\rm qf}$, and
$\sigma_{\rm fast\,fis}$  are the contributions of fusion-fission, quasifission and  fast fission processes, respectively, and $\sigma_{\rm ER}$ is the ER contribution. According to the statement of the authors
of Ref. \cite{ShidlingPLB}, the complete fusion cross sections are obtained by adding fission cross sections \cite{Forster} to the measured data of the evaporation residue cross sections\cite{HindeNPA385}.
Therefore, we can state that the definition of the experimental fusion cross section is similar with the definition of capture \cite{prc792009}: $\sigma^{\rm(exp)}_{\rm fus}=\sigma_{\rm cap}$.

In Ref. \cite{Forster}, the complete fusion cross section is derived from a statistical model where only neutron evaporation and fission are included.
We think that the  fission data from Ref. \cite{Forster} contain quasifission fragments  and, at larger beam energies, also fast fission contributions, which appear as hindrances to  complete fusion.
This argument is confirmed by our results obtained in the framework of the  DNS model. We calculate the  total ER and fusion-fission excitation functions
in the framework of the advanced statistical model \cite{ArrigoPRC1992,ArrigoPRC1994,SagJPG1998}.

\section{Study for the synthesis of the $^{297}$117 element by the $^{48}$Ca+$^{249}$Bk reaction}

\begin{figure}[h]
\centering \vspace{2.5cm}
\resizebox*{1.0\textwidth}{!}{\includegraphics{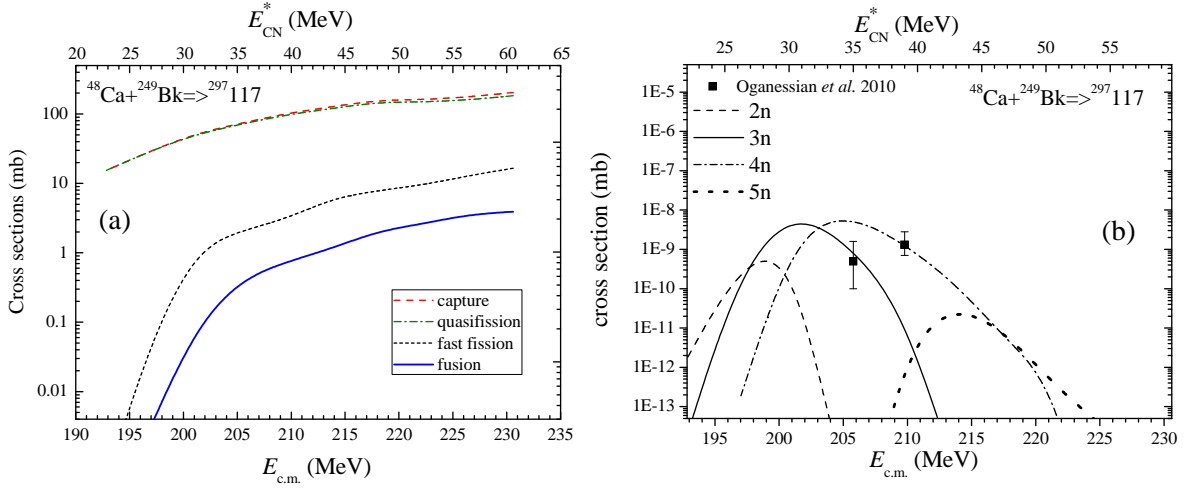}}\vspace{-7.8cm}
\caption{(a) Capture, quasifission, fast fission, and fusion cross sections for the $^{48}$Ca+$^{249}$Bk reaction leading to $^{297}$117 compound nucleus as a function of the colliding energy $E_{\rm c.m.}$.
 (b) Evaporation residue cross sections, after neutron emission only, calculated for the $^{48}$Ca+$^{249}$Bk reaction leading to $^{297}$117 compound nucleus as a function of the colliding energy $E_{\rm c.m.}$,
 in comparison with the experimental data (full square) presented in Ref.\cite{prl2010}.}%\vspace{-0.3cm}
\label{production}
\end{figure}

By using the above-mentioned method we calculated the capture, quasifission, fast fission and fusion (formation of CN) cross sections for the $^{48}$Ca induced reaction on the $^{249}$Bk target (see Fig. \ref{production} (b)).
In the $E_{\rm c.m.}<$200 MeV energy range the fusion cross section is about three orders of magnitude lower than that of quasifission (which is close to the capture cross section in the whole explored energy range),
and it is about one order of magnitude lower than the one of fast fission.
At higher $E_{\rm c.m.}$ energies, the fusion cross section is about two orders of magnitude lower than that of quasifission, while it is about five times lower than the one of fast fission. In this reaction the
quasifission is the completely dominant process in comparison to  complete fusion, and the fast fission  is also a strongly relevant  process in comparison to the fusion formation (CN).
In this case, fast fission process takes place starting from the $\ell_{\rm f}$ value equal to 30 $\hbar$.
\begin{figure}[h]
\centering \vspace{3.5cm}
\includegraphics[width=0.83\textwidth]{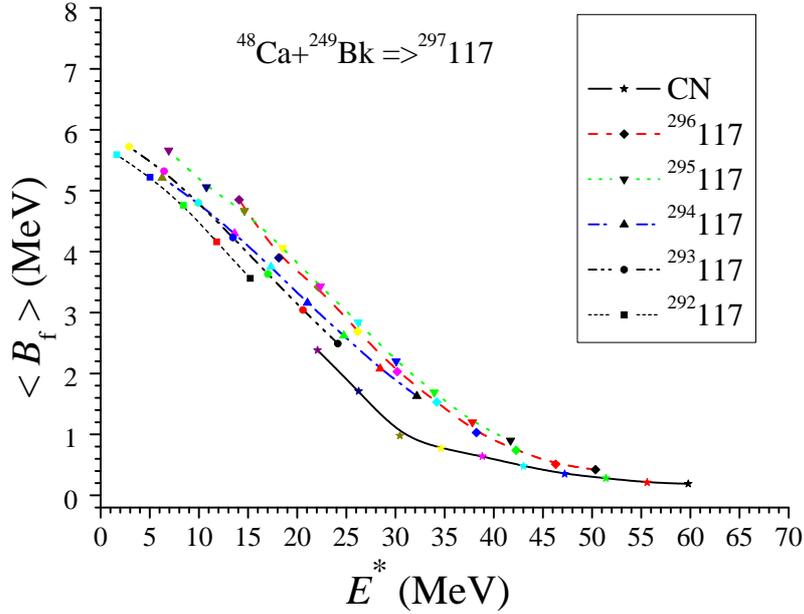}\vspace{-4.5cm}
\caption{The effective fission barrier $<B_{\rm f}>$ as a function
of the excitation energy for the $^{48}$Ca+$^{249}$Bk reaction
leading to $^{297}$117 compound nucleus.}
%\vspace{5.3cm}
\label{figbarr}
\end{figure}

The shell corrections for the odd-odd and odd-even nuclei in the 292-297 mass region of the 117 element were obtained  from the static fission barrier  $B_{\rm f}$ (for $\ell$=0)
calculated by Sobiczewski\cite{privates}, which correspond to the realistic values of shell corrections, as an extension of the results presented in the paper of Kowal {\it et al.} \cite{kowal} for the even-even  superheavy nuclei.
We calculated the average effective fission barriers and  evaporation residue cross sections  for all nuclei along the de-excitation cascade of the $^{297}$117 compound nucleus
 by using such values of shell corrections (for $\ell$=0) and  taking into account  their fade-out  (see Refs. \cite{epja192004,mpla202005,epja222004,jpsj272003} and references therein)
 as a function of the temperature and angular momentum. In  Fig. \ref{production}(b), we compare the individual excitation functions of the evaporation residue for emission of neutron only
  along the cascade of CN with  ER cross sections after  3n and 4n emission from CN measured by Oganessian {\it et al.} in their experiment\cite{prl2010}.
 As one can see our result are in good agreement with the experimental data and we also find that the peaks of such excitation functions for 3n and 4n emission are present at about 30.6 and 34.0 MeV, respectively.
 The maximum yield of the 2n contribution ($5\times10^{-1}$pb) is lower than the ones of the 3n and 4n contributions.
 The maximum yield of the 5n contribution is about 2.3 $\times 10^{-2}$ pb at  $E^*_{CN}$= 43.2 MeV. Moreover,
 in Fig. \ref{figbarr} we report the effective fission barrier $<B_{\rm f}>$ distribution as a function of  $E^*$ along the  cascade of CN at different excitation energies and neutron number of the intermediate excited nuclei.

In Table \ref{tab1} are reported such  $<B_{\rm f}>$ values
obtained for the chain of nuclei, at three different $E*$  values
of the considered excited nuclei.
\begin{table}[h]
%\vspace{1.3cm}
\begin{center}
\caption{Effective fission barrier  $<B_{\rm f}>$ for the
$^{48}$Ca+$^{249}$Bk$\rightarrow ^{297}$117 reaction leading to
intermediate excited nuclei with masses $A$=297, 296, 295, 294,
293 and 292, at excitation energies of nuclei $E^*$= 17, 25 and 33
MeV.}\label{tab1}
\begin{tabular}{|c|c|c|c|c|c|c|}
\hline
$E^*$ (MeV) &  $\begin{array}{c} $A$=297 \\$N$=180 \end{array}$ & $\begin{array}{c} $A$=296 \\$N$=179 \end{array}$& $\begin{array}{c} $A$=295 \\$N$=175 \end{array}$& $\begin{array}{c}
$A$=294 \\$N$=177 \end{array}$ & $\begin{array}{c} $A$=293 \\$N$=176 \end{array}$  & $\begin{array}{c} $A$=292 \\$N$=175 \end{array}$\\
\hline
17 &  .... & 4.2 & 4.3 &  3.8 & 3.6  & 3.3\\
\hline
25 &  1.9 & 2.9 & 3.0 &  2.6 & 2.4  & ....\\
\hline
33 &  0.8 & 1.7 & 1.8 &  1.6 & ....  & ....\\
\hline
\end{tabular}
\end{center}
\end{table}
In  Table \ref{tab1}  the rows show the evolution of the effective fission barrier $<B_{\rm f}>$ values for different neutron number of excited nuclei when they are formed with
a defined excitation energy $E^*$(for example 17, 25, 33 MeV). Of course, such conditions for the intermediate nuclei are reached starting from  different excitation energy values of $E^*_{\rm CN}$.
In details, to form the nuclei with masses $A$ of 297, 296, 295, 294 and 293 u at the same excitation energy, for example $E^*$=25 MeV, it is necessary to start from CN with
excitation energy of 25.0, 33.5, 41.7, 51.6 and 60.8 MeV, respectively.   The changing of the $<B_{\rm f}>$ values along the rows are related to the structure effects of nuclei
 by changing the neutron number. Such changes of $<B_{\rm f}>$  are similar for the three presented cases of  excitation energies.
 Instead, the $<B_{\rm f}>$ values  along a column show the change of the  effective fission barrier for a defined intermediate nucleus by  changing its excitation energy $E^*$.
For all intermediate nuclei with different neutron number, the changes have the same trend by changing the excitation energy $E^*$. It is clear that the $<B_{\rm\, f}>$ value decreases  increasing $E^*$.

\section{Conclusions}

In the large number of  capture reactions with massive nuclei  the
 quasifission process with its peculiarities is the main
      subject for the understanding of the reaction dynamics.
      Therefore, the intense yield of quasifission fragments dominates
in the reactions with  $^{124}$Sn and $^{132}$Sn (neutron-rich)
    beams with targets heavier than  $^{92}$Zr , and
      in the reactions with the Coulomb parameter $z=\frac{Z_1\times Z_2}{A_1^{1/3}+A_2^{1/3}}$  (connected with the intense Coulomb repulsion between  reacting nuclei) higher than 200, for which the
       quasifission  yield is at least two orders of magnitude higher
      than the complete  fusion yield (and then the fusion probability $P_{\rm CN}$ is lower than about 10$^{-2}$, as for example for the reactions listed in Table \ref{tab2}).
\begin{table}[h]
\begin{center}
%\centering
\caption{Reactions, total $Z$ of the system,  $z$ parameter of the
Coulomb repulsion, and  fusion probability $P_{\rm CN}$ are
reported.}\label{tab2}
\begin{tabular}{|c|c|c|c|} \hline
Reaction & $Z_{\rm tot}$ &  $z=\frac{Z_1\times Z_2}{A_1^{1/3}+A_2^{1/3}}$  & $P_{\rm CN}$\\
\hline
$^{86}$Kr+$^{136}$Xe  & 90 &  204  & $\sim$4$\times$10$^{-2}$\\
\hline
$^{92}$Zr+$^{132}$Sn & 90 &  209  & $\sim$5$\times$10$^{-2}$\\
\hline
$^{136}$Xe+$^{136}$Xe& 108 &  284  & $<$10$^{-10}$\\
\hline
$^{64}$Ni+$^{208}$Pb& 110 &  232  & $<$10$^{-5}$\\
\hline
$^{208}$Pb+$^{70}$Ge & 114 &  262  & $<$10$^{-7}$\\
\hline
$^{139,149}$La+ $^{139,149}$La & 114 &  306, 317  & $<$10$^{-10}$\\
\hline
$^{86}$Kr+$^{208}$Pb & 118 &  286  & $<$10$^{-7}$\\
\hline
$^{238}$U+$^{58}$Ni & 120 &  256  & $<$10$^{-6}$\\
\hline
$^{132}$Sn+$^{174}$Yb & 120 &  327  & $<$10$^{-10}$\\
\hline
$^{238}$U+$^{70}$Ge & 124 &  286  & $<$10$^{-7}$\\
\hline
$^{58}$Fe+$^{249}$Cf  & 124 &  251  & $<$10$^{-3}$\\
\hline
$^{132}$Sn+$^{186}$W & 124 &  343  & $<$10$^{-11}$\\
\hline
$^{132}$Sn+$^{208}$Pb & 132 &  373  & $<$10$^{-13}$\\
\hline
$^{160}$Gd+$^{186}$W & 138 &  431  & $<$10$^{-16}$\\
\hline
$^{132}$Sn+$^{249}$Cf  & 148 &  430  & $<$10$^{-15}$\\
\hline
$^{238}$U+$^{248}$Cm  & 188 &  707  & - - - -\\
\hline
\end{tabular}
\end{center}
\hspace{1.5 cm}
\end{table}

       So, all above-mentioned reactions are excellent  reactions for the study of the quasifission process. For the last reaction ($^{238}$U+$^{248}$Cm) in Table \ref{tab2}  the $P_{\rm CN}$
       value is not reported because for this reaction capture cross section does not exist.

      Moreover, at energies higher than the Coulomb barrier, the fast fission process gives a relevant contribution to formation of fissionlike fragments.
For these reactions the contribution of the complete fusion cross section to the capture cross section is very small. Consequently, the  ER cross section is very small too, or in many cases
is lower than the limit (about 0.2 pb) of the present possibility of the experimental setup.  Therefore, it is necessary to estimate the quasifission, fast fission and complete fusion cross sections
by a hopeful analysis of the reaction dynamics before to plan new long beam-time and expensive experiments. Because only the fusion-fission and fusion-evaporation yields contribute to the complete
fusion cross section. Only in the  two above-mentioned reactions ($^{92}$Zr+$^{132}$Sn , $^{86}$Kr+$^{136}$Xe) it is possible to observe evaporation residues. For all other reactions listed in
Table \ref{tab2} it is not possible to observe meaningful fusion-fission contributions and least of all evaporation residues.
The ratio of fusion-fission and  quasifission fragment yields for the all other mentioned reactions are much lower than 10$^{-3}$. Therefore it is impossible to separate the fragments of
fusion-fission process from the fragments of  the huge amount of quasifission process. Our theoretical model is a powerful predictive method to calculate the yield of reaction products and
also to describe the processes in the reactions with massive nuclei.
Moreover, also for reactions used  to synthesis of superheavy elements our model is able to calculate the effective fission barriers along the various steps of the de-excitation cascade
with the neutron and proton emission from  the compound nucleus and intermediate excited nuclear systems  at various excitation energy and angular momentum.

%\newpage

%\subsection{Acknowledgments}
%\ack Authors are  grateful to the Fondazione Bonino-Pulejo of
% Messina for the support received in the collaboration between the Dubna and Messina groups.

%Authors wishing to acknowledge assistance or encouragement from
%colleagues, special work by technical staff or financial support from
%organizations should do so in an unnumbered Acknowledgments section
%immediately following the last numbered section of the paper. The
%command \verb"\ack" sets the acknowledgments heading as an unnumbered
%section.

\end{document}